\documentclass[12pt]{article}

\usepackage[utf8]{inputenc}
\usepackage[T1]{fontenc}
\usepackage{lmodern}
\usepackage[margin=1in]{geometry}
\usepackage{natbib}
\usepackage{amsmath}
\usepackage{booktabs}
\usepackage{longtable}
\usepackage{graphicx}
\usepackage{threeparttable}
\usepackage{siunitx}
\usepackage{float}
\floatstyle{ruled}
\newfloat{algorithm}{t}{loa}
\floatname{algorithm}{Algorithm}
\usepackage{pdfpages}
\usepackage{xr}
\newcommand{\wbb}{\widehat{\boldsymbol \beta}}
\newcommand{\ba}{\boldsymbol \alpha}
\newcommand{\wba}{\widehat{\boldsymbol \alpha}}
\newcommand{\bb}{\boldsymbol \beta}
\newcommand{\uvec}[1]{\mathbf U_{\mathrm{#1}}}

\newcommand{\uveci}{\mathbf U_i}
\newcommand{\uno}{\mathbf U_{\mathrm{G}}}
\newcommand{\lvec}[1]{\boldsymbol \Lambda_{\mathrm{#1}}}

\newcommand{\GEE}{\widehat{\boldsymbol \beta}_{G}}

\newcommand{\GBRR}{\widehat{\boldsymbol \beta}_{RBR}}
\newcommand{\GBRN}{\widehat{\boldsymbol \beta}_{NBR}}
\newcommand{\GBRE}{\widehat{\boldsymbol \beta}_{EBR}}
\newcommand{\GBCR}{\widehat{\boldsymbol \beta}_{RBC}}
\newcommand{\GBCN}{\widehat{\boldsymbol \beta}_{NBC}}
\newcommand{\GBCE}{\widehat{\boldsymbol \beta}_{EBC}}
\newcommand{\mYi}{\mathbf Y_{i}}
\newcommand{\mDi}{\mathbf D_{i}}
\newcommand{\mVi}{\mathbf V_i}
\newcommand{\mVii}{\mathbf V^{-1}_{i}}
\newcommand{\mRi}{\mathbf R_{i}}
\newcommand{\mSi}{\mathbf S_{i}}
\newcommand{\mDeltai}{\boldsymbol \Delta_{i}}
\newcommand{\mAi}{\mathbf A_{i}}
\newcommand{\mSigma}[1]{\boldsymbol \Sigma_{#1}}
\newcommand{\wmSigma}[1]{\widehat{\boldsymbol \Sigma}_{#1}}
\newcommand{\covYi}{\operatorname{cov}\left( \mYi \right)}
\newcommand{\mXi}{\mathbf X_{i}}
\newcommand{\mxij}{\mathbf x_{ij}}
\newcommand{\bmui}{\boldsymbol \mu_{i}}
\newcommand{\wbmui}{\widehat{\boldsymbol \mu}_{i}}
\newcommand{\bigOh}[1]{O \left( {#1} \right)}
\newcommand{\anyvec}[1]{\mathbf {#1}}
\newcommand{\sumin}{\sum_{i=1}^{N}}
\newcommand{\wphi}{\widehat{\phi}}
\newcommand{\identityp}{\mathbf I_{p}}

\newcommand{\onesp}{\mathbf 1_{p}}

\newcommand{\mKi}{\mathbf K_{n_{i}}}

\newcommand{\mvec}[1]{\operatorname{vec}\left( {#1} \right)}

\newcommand{\mderiv}{\mathcal{D}}
\usepackage{hyperref}
\DeclareUnicodeCharacter{2212}{\ensuremath{-}}
\DeclareUnicodeCharacter{2013}{--}
\DeclareUnicodeCharacter{2014}{--}
\DeclareUnicodeCharacter{00A0}{ }

\title{Bias-Reduced GEE via Adjusted Estimating Equations, with Odds-Ratio Extensions}
\author{Anestis Touloumis\\School of Architecture, Engineering and Technology\\University of Brighton\\Brighton, UK\\\texttt{A.Touloumis@brighton.ac.uk}}
\date{}

\begin{document}

\maketitle

\begin{abstract}
Generalized estimating equations (GEE) are widely used for correlated data, but with small to moderate numbers of independent clusters the ordinary GEE regression estimators can be substantially biased. We develop a first-order bias-reduction principle for GEE by viewing the estimator as a clustered-data $M$-estimator and deriving an adjustment to the estimating equations that targets the leading bias term while accounting for the dependence of the working covariance on the mean parameters. The resulting class includes three bias-reduced estimators and three one-step bias-corrected analogs, nesting the bias-corrected estimator of \citet{Lunardon2017} and the bias-reduced and bias-corrected estimators of \citet{Paul_Small_2014} as special cases. The framework applies to general response types through correlation-coefficient parameterizations for the association structure and extends to correlated binary data through pairwise odds-ratio parameterizations, yielding the first bias-reduced and bias-corrected GEE estimators under this parameterization, for which the marginal-mean compatibility constraints are far less restrictive than those of correlation-coefficient parameterizations, making them better suited for small-sample settings. Under standard regularity conditions, all six estimators share the same asymptotic distribution as the ordinary GEE. Simulation studies show that the proposed estimators reduce bias while maintaining efficiency and coverage close to those of ordinary GEE across a range of settings, and a clinical trial analysis illustrates the proposed estimators in practice. Software is available in the \textsf{R} package \texttt{geer}.
\end{abstract}


\section{Introduction}\label{intro}

Generalized estimating equations (GEE), introduced by \citet{Liang1986}, are a standard approach for the analysis of longitudinal and other clustered data. In the GEE framework, the marginal mean is specified through a generalized linear model, while within-cluster dependence is represented through a low-dimensional working association structure. Under mild regularity conditions, GEE yields consistent and asymptotically normal estimators of the regression parameters with valid large-sample standard errors, even when the working association is misspecified. The finite-sample behavior of GEE estimators has received less systematic study, despite growing evidence that bias can be substantial when the number of clusters is small or moderate, which affects Wald-type inference \citep{Paul_Small_2014,Mondol2019,Geroldinger2022,gosho2023}.

Finite-sample bias has long been recognized in likelihood-based inference, and a large literature exists on bias reduction and correction for maximum likelihood estimators, including \citet{cox1968}, \citet{cordeiro1991}, \citet{firth1993}, and \citet{cordeiro1994}. Extending these ideas to GEE is appealing but nontrivial because GEE does not arise from a genuine likelihood and the association structure may be misspecified. \citet{Paul_Small_2014} adapted the approaches of \citet{cox1968} and \citet{firth1993} to obtain bias-corrected and bias-reduced GEE estimators, respectively. \citet{Lunardon2017} noted that their construction relies on correct specification of the association structure and derived a one-step bias-corrected estimator that remains valid under misspecification. A further limitation, not previously noted, is that treating the working covariance matrices as fixed with respect to the marginal means introduces an additional problem even under correct specification of the association structure. In particular, the bias-reduced estimators of \citet{Paul_Small_2014} can fail to achieve first-order bias reduction whenever the variance depends on the mean, as it does for binary and Poisson responses. Our naive bias-reduced and bias-corrected estimators subsume those of \citet{Paul_Small_2014} as special cases and clarify precisely when first-order bias reduction holds. The one-step estimator of \citet{Lunardon2017} arises as a special case of our robust construction. Because all existing bias-corrected estimators require a well-defined ordinary GEE estimator as a starting point, they can encounter numerical difficulties in challenging settings such as separation or near-separation in binary data \citep{albert1984}, where the ordinary GEE estimator may be numerically unstable or difficult to obtain.

For binary responses, a further limitation of existing work is that all bias-correction and bias-reduction methods have focused on correlation-coefficient parameterizations of the association. Feasible correlations are constrained by the marginal means \citep{Prentice_Correlated_1988,Chaganty2004,Chaganty2006}, and in small to moderate samples these constraints can cause correlation estimates to fall outside the admissible range or the working correlation matrix to become nonpositive definite, precisely in the settings where bias reduction is most needed. Pairwise odds-ratio parameterizations \citep{LIPSITZ_Generalized_1991,CAREY_Modelling_1993} avoid these constraints and offer a more stable alternative for modeling association in marginal binary regression, yet no bias-reduction or bias-correction methodology for GEE has been developed under this parameterization, a gap the present paper addresses.

We address these gaps by developing a unified first-order bias-reduction framework for GEE that subsumes and extends existing bias-correction and bias-reduction methods as special cases, embedding the GEE estimator in the clustered-data $M$-estimation framework and building on the robust bias-reduction methodology of \citet{Kosmidis2020}. We derive an explicit target adjustment to the GEE estimating equations that accounts for the dependence of the working covariance matrices on the mean parameters and, from this target, construct three implementable bias-reduced estimators and their one-step bias-corrected analogs. The robust variant replaces unknown covariance matrices with unbiased estimators and achieves first-order bias reduction regardless of whether the working association is correctly specified. Under correct specification, the naive variant uses the model-based working covariance and can offer efficiency gains relative to the robust variant, while the empirical variant replaces expectations with sample analogs and shares the robustness of the robust variant. Under standard regularity conditions, all six estimators are first-order equivalent in distribution to ordinary GEE. The one-step bias-corrected estimators provide a computationally convenient alternative that avoids iterating the adjusted estimating equations. Under correlation-coefficient parameterizations, the robust one-step estimator coincides with the estimator of \citet{Lunardon2017}, and the naive estimators nest those of \citet{Paul_Small_2014} as special cases in which the working covariance matrices are treated as fixed with respect to the mean parameters.

We further extend the framework to odds-ratio parameterizations for correlated binary data, yielding the first bias-reduced and bias-corrected GEE estimators under this parameterization. In challenging settings with correlated binary responses, the proposed estimators may encounter numerical difficulties, and the companion paper by \citet{Touloumis2026} addresses this setting through a penalized GEE framework that provides existence guarantees. An open-source implementation is available in the \textsf{R} package \texttt{geer}.

The remainder of the paper is organized as follows. Section \ref{theory} presents the theoretical developments, derives the ideal adjustment, and introduces the robust, naive, and empirical bias-reduced estimators together with their bias-corrected variants under both correlation-coefficient and odds-ratio parameterizations. Section \ref{sec:simulation} reports simulation studies for correlated binary and Poisson responses that compare competing estimators in terms of bias, efficiency, and coverage. Section \ref{sec:data} illustrates the methodology using a clinical trial application. Section \ref{sec:discussion} summarizes the findings and discusses directions for future research. Derivations, implementation details, and extended numerical results for the simulation study and the application are provided in the Supporting Information.

\section{Bias Reduction via Adjusted Estimating Equations}\label{theory}

\subsection{GEE Setup}\label{theory-gee}

Consider a longitudinal study with $N$ independent subjects. For subject $i$ ($i = 1, \ldots, N$), let $Y_{ij}$ denote the response at time $j$ ($j = 1, \ldots, n_i$), with mean $\mu_{ij}=E(Y_{ij}\mid\mxij)$, where $\mxij$ is a $p$-variate covariate vector. Define $\mYi=(Y_{i1},\ldots,Y_{in_i})^\top$ with mean vector $\bmui$, and let $\mXi$ be the corresponding $n_i\times p$ covariate matrix. We assume the marginal regression model
\[
g(\mu_{ij})=\eta_{ij}=\mxij^\top\bb,
\]
where $g(\cdot)$ is a known link function and $\bb$ is the regression parameter vector.

Let $\operatorname{Var}(Y_{ij})=\phi h(\mu_{ij})$, where $h(\cdot)$ is a known variance function and $\phi$ is a dispersion parameter. Write $\mAi$ for the $n_i \times n_i$ diagonal matrix with $j$th diagonal element $h(\mu_{ij})$, and let $\mRi(\ba)$ be the $n_i \times n_i$ working correlation matrix depending on a nuisance parameter vector $\ba$. The working covariance matrix is $\mVi = \phi\,\mAi^{1/2}\mRi(\ba)\mAi^{1/2}$. Given $\phi$ and $\ba$, \citet{Liang1986} defined the GEE estimator $\GEE$ as
the solution of
\[
\uvec{G} = \sumin \uveci = \sumin \mDi^{\top} \mVii \mSi = \mathbf 0_p,
\]
where $\mSi = \mYi - \bmui$ and $\mDi = \mDeltai\mXi$, with $\mDeltai$ the $n_i \times n_i$ diagonal matrix with $j$th diagonal element $\partial\mu_{ij}/\partial\eta_{ij}$. When $\ba$ and $\phi$ are unknown, they are replaced by $\sqrt{N}$-consistent estimators $\wba$ and $\wphi$, typically obtained by the method of moments \citep{Liang1986}.

Under standard regularity conditions, $\GEE$ is consistent and asymptotically normal with covariance matrix $\mSigma{G}=\mSigma{0}^{-1}\mSigma{1}\mSigma{0}^{-1}$, where $\mSigma{0}=\sumin\mDi^{\top}\mVii\mDi$ and $\mSigma{1}=\sumin E(\uveci\uveci^{\top}) = \sumin \mDi^{\top} \mVii \covYi \mVii \mDi$. This is consistently estimated by the sandwich estimator
\begin{equation}
\wmSigma{R} = \wmSigma{0}^{-1}\wmSigma{1}\wmSigma{0}^{-1},
\label{sigma-r}
\end{equation}
where $\wmSigma{0}$ and $\wmSigma{1}$ are obtained by replacing ($\bb$,$\ba$, $\phi$) with ($\GEE$,$\wba$,$\wphi$) and $\wmSigma{1}$ additionally replaces $\covYi$ by $(\mYi-\wbmui)(\mYi-\wbmui)^{\top}$. In small samples, $\wmSigma{R}$ can underestimate sampling variability and inflate type I error. We therefore use the small-sample adjusted estimator of \citet{Morel_Small_2003},
\begin{equation}
\wmSigma{M} = \frac{n^{\star}-1}{n^{\star}-p}\frac{N}{N-1}\wmSigma{R}
+ \lambda\xi\wmSigma{0}^{-1},
\label{sigma-m}
\end{equation}
where $n^{\star}=\sumin n_i$, $\lambda=\min\{0.5,p/(N-p)\}$, and $\xi=\max\{1,\operatorname{tr}(\wmSigma{0}^{-1}\wmSigma{1})/p\}$. The estimator $\wmSigma{M}$ remains positive definite whenever $\wmSigma{0}$ is positive definite.

For vector and matrix derivatives, we adopt the narrow definition of \citet{Magnus2010}. If $\mathbf a=(a_1,\ldots,a_c)^\top$ is a vector function of $\mathbf b=(b_1,\ldots,b_d)^\top$, then $\partial \mathbf a/\partial \mathbf b^\top$ denotes the $c\times d$ matrix with $(i,j)$ entry $\partial a_i/\partial b_j$. If $\mathbf A$ is a matrix function of a vector $\mathbf b$, then $\mderiv \mathbf A (\mathbf b)=\partial\, \mvec{\mathbf A}/\partial \mathbf b^\top$.

\subsection{The Bias-Reducing Adjustment and Its Target}\label{sec:br-framework}
The derivations assume the $M$-estimation regularity conditions of \citet{Kosmidis2020} and the assumptions of Theorem 2 of \citet{Liang1986}.

Treating $\ba$ and $\phi$ as known, $\GEE$ is an $M$-estimator with bias expansion $E(\GEE - \bb) = \mSigma{0}^{-1} \lvec{G} + O(N^{-3/2})$, where
\begin{equation}
  \lvec{G}
  =
  -
  \left( \identityp \otimes \onesp^{\top} \right)
  \left[
    \left( \onesp \otimes \mSigma{0}^{-1} \right) \circ \lvec{G1}
    +
    \tfrac{1}{2}
    \left( \onesp \otimes \mSigma{G} \right) \circ \lvec{G2}
  \right]
  \onesp,
\label{eq:ideal-adjustment}
\end{equation}
with
\[
\begin{aligned}
\lvec{G1}
& = \sumin \left(\mDi \otimes \mDi \right)^{\top} \lvec{G1i} \covYi \mVii \mDi, \\
\lvec{G1i} &= \mKi \left(\mDeltai^{\star} + \mAi^{\star}\right) \mVii +
  \left(\mVii \otimes \mathbf I_{n_i}\right) \mKi \mAi^{\star}, \\
\lvec{G2} & = - \sumin \left(\mDi \otimes \mDi \right)^{\top}
  \left(\lvec{G1i} + \lvec{G2i} \right) \mDi, \\
\lvec{G2i} &= \left(\mVii \otimes \mVii \right) \mKi
  \left(\mDeltai^{\star} + \mAi^{\star}\right).
\end{aligned}
\]
The definitions of $\mKi$, $\mDeltai^{\star}$ and $\mAi^{\star}$, and a detailed derivation of \eqref{eq:ideal-adjustment} are given in Supporting Information, Web Appendix B. We treat $\lvec{G}$ as the ideal target and construct practical adjustments $\lvec{BR}$ that approximate it. Following \citet{Kosmidis2020}, we define bias-reduced estimators as solutions of
\begin{equation}
  \uvec{BR} = \uno + \lvec{BR} = \mathbf 0_p,
  \label{adjusted-equation-gee}
\end{equation}
where $\lvec{BR}$ depends on $\bb$, $\ba$, $\phi$ and the data, and its derivatives with respect to $(\bb, \ba, \phi)$ are $O_p(1)$. When
condition
\begin{equation}
  E(\lvec{BR}) = \lvec{G} + O(N^{-1/2})
  \label{adjustment-vector-expectation}
\end{equation}
holds, $E(\wbb_{BR} - \bb) = O(N^{-3/2})$ and $N^{1/2}(\wbb_{BR} - \bb) \overset{d}{\longrightarrow} \mathcal{N}_p\!\left(\mathbf{0}_p, \lim_{N\to\infty} N\mSigma{G}\right)$. A proof sketch of asymptotic normality is given in Web Appendix E of the Supporting Information.

The corresponding one-step bias-corrected estimator,
\begin{equation}
  \wbb_{BC} = \GEE - \mSigma{0}^{-1}\!\bigl(\GEE\bigr)\,
  \lvec{BR}\!\bigl(\GEE\bigr),
  \label{equation-bcgee}
\end{equation}
has bias of order $o(N^{-1})$ \citep{Kosmidis2020} and the same asymptotic distribution. Both $\wbb_{BR}$ and $\wbb_{BC}$ may use $\wmSigma{R}$ or $\wmSigma{M}$ for inference, with $\GEE$ replaced by $\wbb_{BR}$ or $\wbb_{BC}$ as appropriate. When nuisance parameters are unknown, we follow \citet{Paul_Small_2014} and \citet{Lunardon2017} and replace $(\ba,\phi)$ by $\sqrt{N}$-consistent estimators $(\wba,\wphi)$, either fixed at the ordinary GEE values or updated jointly with $\bb$ at each iteration, as described in Algorithm \ref{alg:brgee}.

\begin{algorithm}[t]
\caption{Computation of bias-reduced and bias-corrected GEE estimators}
\label{alg:brgee}
\smallskip
\noindent\textbf{Input:} clustered data $\{(\mathbf{Y}_i, \mathbf{X}_i)\}_{i=1}^N$; link $g$; variance $h$; working association model; bias-reduction or bias-correction method; nuisance update option; tolerance $\varepsilon_\beta$; maximum iterations $B$\\[2pt]
\noindent\textbf{Output:} $\widehat{\boldsymbol\beta}$ and standard errors via \eqref{sigma-r} or \eqref{sigma-m}
\smallskip\hrule\smallskip
\noindent 1.\enspace Initialize $\boldsymbol\beta^{(0)}$ by fitting a GLM under working independence. If initialization fails (e.g., due to separation), reinitialize $\boldsymbol\beta^{(0)}$ using Firth's bias-reduced GLM.\\[2pt]
\noindent 2.\enspace Initialize nuisance $(\boldsymbol\alpha^{(0)}, \phi^{(0)})$. For odds-ratio parameterizations, set $\phi^{(0)} = 1$.\\[4pt]
\noindent\textbf{If} the method is one-step (RBC/NBC/EBC):\\[2pt]
\hspace*{2em} Compute $(\widehat{\boldsymbol\beta}_G, \widehat{\boldsymbol\alpha}, \widehat\phi)$ and return
$\widehat{\boldsymbol\beta} = \widehat{\boldsymbol\beta}_G -
\boldsymbol\Sigma_0(\widehat{\boldsymbol\beta}_G, \widehat{\boldsymbol\alpha}, \widehat\phi)^{-1}
\boldsymbol\Lambda_{\mathrm{BR}}(\widehat{\boldsymbol\beta}_G, \widehat{\boldsymbol\alpha}, \widehat\phi)$.\\[4pt]
\noindent\textbf{For} $b = 0, 1, \ldots, B-1$:\\[2pt]
\hspace*{2em} (a)\enspace Update nuisance $(\boldsymbol\alpha^{(b)}, \phi^{(b)})$ if joint updates are used.\\[2pt]
\hspace*{2em} (b)\enspace Compute
$\mathbf{U}_{\mathrm{BR}}(\boldsymbol\beta^{(b)}) =
  \mathbf{U}_{\mathrm{G}}(\boldsymbol\beta^{(b)}, \boldsymbol\alpha^{(b)}, \phi^{(b)}) +
  \boldsymbol\Lambda_{\mathrm{BR}}(\boldsymbol\beta^{(b)}, \boldsymbol\alpha^{(b)}, \phi^{(b)})$.\\[2pt]
\hspace*{2em} (c)\enspace Set
$\boldsymbol\beta^{(b+1)} = \boldsymbol\beta^{(b)} -
  \boldsymbol\Sigma_0(\boldsymbol\beta^{(b)}, \boldsymbol\alpha^{(b)}, \phi^{(b)})^{-1}
  \mathbf{U}_{\mathrm{BR}}(\boldsymbol\beta^{(b)})$.\\[2pt]
\hspace*{2em} (d)\enspace \textbf{If} $\|\boldsymbol\beta^{(b+1)} - \boldsymbol\beta^{(b)}\|_\infty < \varepsilon_\beta$,
  return $\widehat{\boldsymbol\beta} = \boldsymbol\beta^{(b+1)}$.\\[4pt]
\noindent Report nonconvergence and return last iterate.
\smallskip\hrule
\end{algorithm}

\subsection{Three Approximations to the Ideal Adjustment}\label{sec:approx}

The ideal target $\lvec{G}$ in \eqref{eq:ideal-adjustment} depends on the unknown covariance matrices $\{\covYi: i=1, \ldots, N\}$. We consider three implementable adjustment vectors satisfying \eqref{adjustment-vector-expectation}: a robust adjustment based on unbiased covariance estimates, a naive adjustment that requires correct specification of the association structure, and an empirical adjustment based on sample analogs throughout.

\textbf{Robust adjustment.} Replacing each $\covYi$ by the unbiased analog $\mSi\mSi^{\top}$ gives
\begin{equation}
\lvec{R}
=
-
\left(\identityp\otimes \onesp^{\top}\right)
\left\{
\left[\left(\onesp\otimes \mSigma{0}^{-1}\right)\circ \lvec{R1}\right]
+
\tfrac{1}{2}
\left[\onesp\otimes \left(\mSigma{0}^{-1}\anyvec{e}\,\mSigma{0}^{-1}\right)\right]
\circ \lvec{G2}
\right\}\onesp,
\label{adjustment-robust}
\end{equation}
where $\lvec{R1}=\sumin(\mDi\otimes \mDi)^{\top}\lvec{G1i}\,\mSi\mSi^{\top}\mVii\mDi$ and $\anyvec{e}=\sumin \uveci \uveci^{\top}$ is the empirical analog of $\mSigma{1}$. Because $E(\mSi\mSi^{\top})=\covYi$, we have $E(\lvec{R})=\lvec{G}$ exactly, so \eqref{adjustment-vector-expectation} holds. The robust bias-reduced (RBR) estimator $\GBRR$ solves
\[
\uvec{R}=\uno+\lvec{R}=\mathbf 0_p
\]
and does not require correct specification of the working association structure.

\textbf{Naive adjustment.} Assuming $\mVi=\covYi$ for all $i$, the target
$\lvec{G}$ simplifies to
\begin{equation}
\lvec{N}
=
  -
  \left( \identityp \otimes \onesp^{\top} \right)
\left[
  \left( \onesp \otimes \mSigma{0}^{-1} \right) \circ
  \left( \lvec{N1} + \tfrac{1}{2} \lvec{N2} \right)
  \right]
\onesp,
\label{adjustment-naive}
\end{equation}
where
\[
  \lvec{N1}
   = \tfrac{1}{2}
  \sumin \left( \mDi \otimes \mDi \right)^{\top}
         \left[ \mKi \mDeltai^{\star} \mVii -
         \left( \mathbf I_{n_i} \otimes \mVii \right) \mKi \mDeltai^{\star}
         -
         \left( \mVii \otimes \mathbf I_{n_i} \right)
         \mKi \mDeltai^{\star} \right] \mDi
\]
arises from derivatives of the mean component, and
\[
  \lvec{N2}
   =
  \sumin \left( \mDi \otimes \mDi \right)^{\top}
  \left[ \mKi \mAi^{\star} \mVii
         - 2\left( \mathbf I_{n_i} \otimes \mVii \right) \mKi \mAi^{\star}
         +
         \left( \mVii \otimes \mathbf I_{n_i} \right)\mKi \mAi^{\star}
    \right]
   \mDi
\]
captures the dependence of the variance function $h(\bmui)$ on $\bb$. The naive bias-reduced (NBR) estimator $\GBRN$ solves $\uvec{N} = \uno + \lvec{N} = \mathbf 0_p$. Under misspecification of the association structure, first-order bias reduction holds only when
\begin{equation}
\lvec{G} - \lvec{N} = \bigOh{N^{-1/2}}.
\label{naive_condition}
\end{equation}
Supporting Information, Web Appendix D provides an example in which condition \eqref{naive_condition} holds under misspecification. This implies that NBR can achieve bias reduction despite incorrect working association specification, a pattern confirmed by the simulation study in Section \ref{sec:simulation}.

Setting $\lvec{N2}=\mathbf 0_p$ recovers the estimator of \citet{Paul_Small_2014}, which treats $\mVi$ as fixed with respect to $\bb$ and does not in general deliver first-order bias reduction for mean-dependent variance functions, regardless of whether the working association is correctly specified. Under working independence, $\GBRN$ reduces to the bias-reduced GLM estimator of \citet{firth1993}.

\textbf{Empirical adjustment.}  Let $\anyvec{j} = \mderiv\uno(\bb)$, $\anyvec{d} = \sumin \operatorname{vec}\!\left(\left[\mderiv\uveci(\bb) \right]^{\top}\right)\uveci^{\top}$, and $\anyvec{u} =
\left[\mderiv\anyvec{j}(\bb)\right]^{\top}$ be empirical analogs of $\mSigma{0}$, $\lvec{G1}$ and $\lvec{G2}$ (see Web Appendix B of the Supporting Information). The empirical adjustment is
\begin{equation}
  \lvec{E}
  =
  -
  \left(\identityp \otimes \onesp^{\top}\right)
  \left[
    \left(\onesp \otimes \anyvec{j}^{-\top}\right) \circ \anyvec{d}
    +
    \tfrac{1}{2}
    \left(\onesp \otimes \anyvec{j}^{-1}\anyvec{e}\,\anyvec{j}^{-\top}\right)
    \circ \anyvec{u}
  \right]
  \onesp,
\label{adjustment_empirical}
\end{equation}
and the empirical bias-reduced (EBR) estimator $\GBRE$ solves $\uvec{E} = \uno + \lvec{E} = \mathbf 0_p$. Although $\lvec{E}$ satisfies \eqref{adjustment-vector-expectation} under the conditions of \citet{Kosmidis2020}, it can differ from $\lvec{R}$ by zero-mean terms that can inflate finite-sample variability. See Web Appendix D of the Supporting Information for an example. We therefore recommend $\GBRR$ as the default and treat $\GBRE$ as a fallback when the robust procedure is numerically unstable.

\subsection{Binary Responses: Odds-Ratio Parameterization}\label{sec:or}
For correlated binary responses, pairwise odds-ratio parameterizations \citep{LIPSITZ_Generalized_1991,CAREY_Modelling_1993} avoid the restrictive feasibility constraints on correlation coefficients \citep{Chaganty2004,Chaganty2006}, making them better suited for measuring the association of correlated binary responses. The dispersion parameter is fixed at $\phi=1$, the nuisance vector $\ba$ collects pairwise odds ratios from within-subject $2\times2$ tables, and the working covariance $\mVi$ is specified through $\mu_{ijj^{\prime}} = \Pr(Y_{ij}=Y_{ij^{\prime}}=1)$, with the $(j,j^{\prime})$th off-diagonal element of $\mVi$ equal to $\mu_{ijj^{\prime}}-\mu_{ij}\mu_{ij^{\prime}}$. For subject $i$ and times $j \neq j^{\prime}$,
\[
\mu_{ijj^{\prime}}=
\begin{cases}
\dfrac{f_{ijj^{\prime}}-\sqrt{f^2_{ijj^{\prime}}-4\alpha_{ijj^{\prime}}(\alpha_{ijj^{\prime}}-1)\mu_{ij}\mu_{ij^{\prime}}}}{2(\alpha_{ijj^{\prime}}-1)} & \mbox{if $\alpha_{ijj^{\prime}}\neq 1$}, \\[6pt]
\mu_{ij}\mu_{ij^{\prime}} & \mbox{if $\alpha_{ijj^{\prime}}=1$},
\end{cases}
\]
where $f_{ijj^{\prime}}=1-(1-\alpha_{ijj^{\prime}})(\mu_{ij}+\mu_{ij^{\prime}})$ and $\alpha_{ijj^{\prime}}$ is the odds ratio for the pair $(j,j^{\prime})$. When $\ba$ is unknown, it can be estimated via additional estimating equations \citep{LIPSITZ_Generalized_1991,CAREY_Modelling_1993} or via the marginalized odds-ratio approach of \citet{Touloumis2013}, as implemented in \citet{Touloumis2026} for correlated binary responses.

The bias-reduction framework of Section \ref{sec:br-framework} applies unchanged, but $\lvec{G1}$ and $\lvec{G2}$ must be rederived because $\mVi$ no longer decomposes as $\mAi^{1/2}\mRi(\ba)\mAi^{1/2}$. The odds-ratio counterparts $\lvec{G1}^{OR}$ and $\lvec{G2}^{OR}$ are given in Supporting Information, Web Appendix C. The robust, naive and empirical adjustments under odds ratios are obtained from \eqref{adjustment-robust}, \eqref{adjustment-naive} and \eqref{adjustment_empirical}, respectively, by replacing the correlation-based components with their odds-ratio counterparts.

\subsection{Bias Correction}\label{sec:bc}

Using \eqref{equation-bcgee}, the robust, naive and empirical bias-corrected estimators under correlation-coefficient
parameterizations are
\[
\begin{aligned}
\GBCR &= \GEE - \mSigma{0}^{-1}\!\bigl(\GEE\bigr)\,\lvec{R}\!\bigl(\GEE\bigr), \\
\GBCN &= \GEE - \mSigma{0}^{-1}\!\bigl(\GEE\bigr)\,\lvec{N}\!\bigl(\GEE\bigr), \\
\GBCE &= \GEE - \anyvec{j}^{-1}\!\bigl(\GEE\bigr)\,\lvec{E}\!\bigl(\GEE\bigr)
\end{aligned}
\]
where $\anyvec{j}(\GEE)$ replaces $\mSigma{0}(\GEE)$ in the empirical estimator, consistent with the approach of \citet{Kosmidis2020}. The odds-ratio counterparts are obtained by replacing each adjustment vector with its odds-ratio version. When $\GEE$ is unavailable, the bias-corrected estimators cannot be formed, and the fully iterated estimators should be initialized as described in Algorithm \ref{alg:brgee}. Under the correlation-coefficient parameterization, $\GBCR$ coincides with the robust bias-corrected estimator of \citet{Lunardon2017}, and $\GBCN$ recovers the estimator of \citet{Paul_Small_2014} when $\lvec{N2}=\mathbf 0_p$.

\subsection{Limitations and Practical Guidance}\label{sec:practical}
Three limitations are worth noting. First, first-order bias reduction is not guaranteed for the naive estimators NBR and NBC unless the association structure has been modeled correctly or condition \eqref{naive_condition} holds. Second, for correlated binary responses the odds-ratio parameterization of Section \ref{sec:or} should be preferred, because under correlation-coefficient parameterizations, the working correlation matrix may violate the Fréchet bounds and lead to inadmissible values of $\bb$ and $\ba$ as noted by \citet{Prentice_Correlated_1988}, \citet{Chaganty2004}, and \citet{Chaganty2006}. Third, when the ordinary GEE solution $\GEE$ does not exist, the bias-corrected estimators RBC, NBC and EBC are unavailable. In such cases the fully iterated bias-reduced estimators should be initialized as described in Algorithm \ref{alg:brgee}, and for binary responses the penalized GEE framework of \citet{Touloumis2026} provides a complementary approach with existence guarantees.

For routine applications we recommend RBR with the small-sample covariance adjustment of \citet{Morel_Small_2003} and an odds-ratio parameterization for correlated binary responses. NBR may be useful when the working association model is believed to be close to correct, while EBR is mainly a fallback when RBR is numerically unstable. Fully iterated estimators are preferred when feasible, and one-step corrections are useful when $\GEE$ exists and the iterative algorithm converges rapidly.

\subsection{Software}\label{sec:software}
All proposed estimators are implemented in the \textsf{R} package \texttt{geer}, available from CRAN. The function \texttt{geewa()} fits marginal models under correlation-coefficient parameterizations, while \texttt{geewa\_binary()} fits correlated binary-response models under odds-ratio parameterizations, with odds ratios estimated via the marginalized approach of \citet{Touloumis2026}. In both functions, the argument \texttt{method} selects ordinary GEE, the fully iterated bias-reduced estimators, or their one-step bias-corrected counterparts. The default is \texttt{method = "gee"}, and users seeking robust bias-reduction or bias-correction should specify \texttt{method = "brgee-robust"} or \texttt{method = "bcgee-robust"}, respectively.

\section{Simulation Study}\label{sec:simulation}
This section reports simulation results assessing the finite-sample performance of bias reduction for GEE with correlated binary and Poisson responses. We compare ordinary GEE with the three proposed fully iterated bias-reduced estimators: RBR, NBR, and EBR. One-step bias-corrected variants are omitted because their simulation results closely track the fully iterated counterparts. We also omit the naive estimators of \citet{Paul_Small_2014} since they correspond to the special case $\lvec{N2}=\mathbf 0_p$ of our naive adjustment (Section \ref{sec:approx}).

For each configuration, we generated 10,000 Monte Carlo replications. For each regression parameter and method, we recorded the Monte Carlo bias ($\times 100$), empirical standard error (ESE), empirical coverage (EC, \%) of nominal $95\%$ Wald confidence intervals based on $\wmSigma{M}$, and the convergence proportion (CP, \%). We also computed two relative efficiency measures. The relative efficiency (RE) is the ratio of the theoretical standard error derived from $\wmSigma{M}$ to the ESE, with values close to 1 indicating that $\wmSigma{M}$ accurately reflects the true sampling variability. The simulated relative efficiency (SRE) is the ratio of the MSE of the ordinary GEE estimator under the working independence model to the MSE of the estimator under consideration, where $\text{MSE} = \text{Bias}^2 + \text{ESE}^2$ with Bias on the original scale, with the GEE estimator under the working independence model serving as a common baseline across all simulation configurations. An estimator was declared to have converged if the iterated estimating equations satisfied a relative tolerance of $10^{-6}$ within at most $500$ iterations, $\max_k|\widehat\beta_k|<10^2$  to flag numerical divergence, and $\wmSigma{M}$ could be formed without numerical warnings. Replications failing any of these criteria were treated as nonconverged.

\subsection{Binary Responses}\label{sec:sim-binary}
Binary outcomes were generated from the probit marginal model
\[
\Phi^{-1}(\pi_{ij})=\beta_0+\beta_1 x_{1i}+\beta_2 x_{2ij},
\]
for clusters $i=1,\ldots,N$ and occasions $j=1,2,3,4$, with $(\beta_0,\beta_1,\beta_2)^{\top}=(0,0.5,1)^{\top}$ and $N\in\{20,35,50,100,500\}$. Covariates were generated as $x_{1i}\stackrel{\mathrm{iid}}{\sim}\mathcal{N}(0,0.5^2)$ and $\mathbf{x}_{2i}\stackrel{\mathrm{iid}}{\sim}\mathcal{N}(\mathbf{0}_4,\mSigma{x})$ with $\mSigma{x}=0.5^2(0.2\mathbf{I}_4+0.8\mathbf{1}_4\mathbf{1}_4^{\top})$. Correlated binary responses were generated via the threshold method in \texttt{SimCorMultRes} \citep{Touloumis2016a} using the latent correlation matrix
\[
\mathbf{R}=
\begin{bmatrix}
1    & 0.85 & 0.50 & 0.15 \\
0.85 & 1    & 0.85 & 0.50 \\
0.50 & 0.85 & 1    & 0.85 \\
0.15 & 0.50 & 0.85 & 1
\end{bmatrix}.
\]
We fitted the marginal model using GEE, RBR, NBR, and EBR under two working odds-ratio structures (independence and exchangeable). Because neither working structure coincides with the dependence induced by
$\mathbf{R}$, these experiments also assess performance under misspecification of the association structure. We also considered an MCAR \citep{rubin1976inference} scenario with monotone dropout: responses were always observed at $j=1$, and each subject independently dropped out at $j=2,3,4$ with probability $0.2$, after which all subsequent responses were missing. Results for the complete and MCAR scenarios are reported in Tables \ref{tab:sim_binary_complete_merged} and \ref{tab:sim_binary_mcar_merged}, respectively.

\begin{table}[H]
\centering
\caption{Binary responses (probit model), complete data. Results for $\beta_1=0.5$
and $\beta_2=1$ under working odds-ratio structures (independence and exchangeable).
Columns report Bias ($100\times$Bias), empirical standard error (ESE),
relative efficiency (RE), simulated relative efficiency (SRE),
empirical coverage (EC, \%) of nominal 95\% Wald intervals,
and convergence proportion (CP, \%).}
\label{tab:sim_binary_complete_merged}
\resizebox{\textwidth}{!}{%
\begin{tabular}{@{}
S[table-format=3.0] l
S[table-format=+3.2] S[table-format=1.4] S[table-format=1.2] S[table-format=1.2]
S[table-format=2.2] S[table-format=3.2]
S[table-format=+3.2] S[table-format=1.4] S[table-format=1.2] S[table-format=1.2]
S[table-format=2.2] S[table-format=3.2]@{}}
\toprule
& & \multicolumn{6}{c}{Independence} & \multicolumn{6}{c}{Exchangeable} \\
\cmidrule(lr){3-8}\cmidrule(lr){9-14}
{$N$} & {Method}
& {Bias} & {ESE} & {RE} & {SRE} & {EC} & {CP}
& {Bias} & {ESE} & {RE} & {SRE} & {EC} & {CP} \\
\midrule

\multicolumn{14}{c}{$\beta_1=0.5$}\\
\addlinespace[3pt]
    20 & GEE & +2.92 & 0.5438 & 0.95 & 1.00 & 93.48 & 100.00 & +3.36 & 0.5402 & 0.97 & 1.01 & 93.96 & 100.00 \\
     & RBR & -0.22 & 0.5032 & 1.00 & 1.17 & 94.07 & 100.00 & +0.28 & 0.5057 & 1.01 & 1.16 & 94.46 & 99.96 \\
     & NBR & +1.08 & 0.5132 & 0.99 & 1.13 & 94.04 & 100.00 & -0.58 & 0.4813 & 1.06 & 1.28 & 95.36 & 100.00 \\
     & EBR & -0.05 & 0.5106 & 0.98 & 1.14 & 93.80 & 100.00 & +0.42 & 0.5098 & 1.00 & 1.14 & 94.35 & 100.00 \\
\addlinespace[3pt]
    35 & GEE & +3.42 & 0.4129 & 0.97 & 1.00 & 94.14 & 100.00 & +2.94 & 0.4107 & 0.97 & 1.01 & 94.01 & 100.00 \\
     & RBR & +0.73 & 0.3931 & 1.00 & 1.11 & 94.66 & 100.00 & +0.61 & 0.3935 & 1.00 & 1.11 & 94.55 & 99.98 \\
     & NBR & +1.79 & 0.4007 & 0.99 & 1.07 & 94.54 & 100.00 & -0.06 & 0.3867 & 1.02 & 1.15 & 94.93 & 100.00 \\
     & EBR & +1.03 & 0.3949 & 1.00 & 1.10 & 94.62 & 100.00 & +0.74 & 0.3947 & 1.00 & 1.10 & 94.54 & 100.00 \\
\addlinespace[3pt]
    50 & GEE & +2.90 & 0.3156 & 1.00 & 1.00 & 94.98 & 100.00 & +2.75 & 0.3141 & 1.00 & 1.01 & 94.91 & 100.00 \\
     & RBR & +1.06 & 0.3047 & 1.02 & 1.08 & 95.19 & 100.00 & +1.07 & 0.3044 & 1.02 & 1.08 & 95.26 & 100.00 \\
     & NBR & +1.92 & 0.3092 & 1.01 & 1.05 & 95.20 & 100.00 & +0.82 & 0.3017 & 1.03 & 1.10 & 95.45 & 100.00 \\
     & EBR & +1.21 & 0.3055 & 1.02 & 1.07 & 95.16 & 100.00 & +1.20 & 0.3049 & 1.02 & 1.08 & 95.28 & 100.00 \\
\addlinespace[3pt]
    100 & GEE & +1.18 & 0.2135 & 0.99 & 1.00 & 94.56 & 100.00 & +1.18 & 0.2133 & 0.99 & 1.00 & 94.51 & 100.00 \\
     & RBR & +0.21 & 0.2093 & 1.00 & 1.04 & 94.80 & 100.00 & +0.25 & 0.2096 & 1.00 & 1.04 & 94.74 & 100.00 \\
     & NBR & +0.70 & 0.2112 & 1.00 & 1.02 & 94.72 & 100.00 & +0.13 & 0.2084 & 1.00 & 1.05 & 94.80 & 100.00 \\
     & EBR & +0.25 & 0.2094 & 1.00 & 1.04 & 94.77 & 100.00 & +0.28 & 0.2097 & 1.00 & 1.04 & 94.73 & 100.00 \\
\addlinespace[3pt]
    500 & GEE & +0.27 & 0.0948 & 1.00 & 1.00 & 94.66 & 100.00 & +0.27 & 0.0948 & 1.00 & 1.00 & 94.60 & 100.00 \\
     & RBR & +0.06 & 0.0944 & 1.00 & 1.01 & 94.67 & 100.00 & +0.06 & 0.0944 & 1.00 & 1.01 & 94.72 & 100.00 \\
     & NBR & +0.17 & 0.0946 & 1.00 & 1.00 & 94.65 & 100.00 & +0.06 & 0.0943 & 1.00 & 1.01 & 94.71 & 100.00 \\
     & EBR & +0.06 & 0.0944 & 1.00 & 1.01 & 94.67 & 100.00 & +0.06 & 0.0944 & 1.00 & 1.01 & 94.72 & 100.00 \\

\addlinespace[6pt]
\multicolumn{14}{c}{$\beta_2=1$}\\
\addlinespace[3pt]
    20 & GEE & +11.79 & 0.5594 & 0.94 & 1.00 & 93.26 & 100.00 & +8.01 & 0.4863 & 0.95 & 1.35 & 93.65 & 100.00 \\
     & RBR & +3.38 & 0.5147 & 0.97 & 1.23 & 93.66 & 100.00 & +0.56 & 0.4539 & 0.97 & 1.59 & 93.93 & 99.96 \\
     & NBR & +6.04 & 0.5198 & 0.97 & 1.19 & 93.75 & 100.00 & -2.04 & 0.4272 & 1.03 & 1.79 & 94.49 & 100.00 \\
     & EBR & +4.79 & 0.5305 & 0.94 & 1.15 & 93.22 & 100.00 & +1.37 & 0.4591 & 0.96 & 1.55 & 93.77 & 100.00 \\
\addlinespace[3pt]
    35 & GEE & +6.65 & 0.3990 & 0.97 & 1.00 & 94.01 & 100.00 & +4.21 & 0.3442 & 0.98 & 1.36 & 94.14 & 100.00 \\
     & RBR & +1.09 & 0.3798 & 0.98 & 1.13 & 94.41 & 100.00 & -0.42 & 0.3301 & 0.99 & 1.50 & 94.23 & 99.98 \\
     & NBR & +3.20 & 0.3842 & 0.99 & 1.10 & 94.41 & 100.00 & -1.63 & 0.3220 & 1.02 & 1.57 & 94.53 & 100.00 \\
     & EBR & +1.79 & 0.3854 & 0.97 & 1.10 & 94.05 & 100.00 & -0.08 & 0.3310 & 0.99 & 1.49 & 94.20 & 100.00 \\
\addlinespace[3pt]
    50 & GEE & +4.40 & 0.3441 & 0.99 & 1.00 & 94.31 & 100.00 & +2.61 & 0.2962 & 0.98 & 1.36 & 94.54 & 100.00 \\
     & RBR & +0.58 & 0.3317 & 1.00 & 1.09 & 94.81 & 100.00 & -0.70 & 0.2866 & 1.00 & 1.46 & 94.64 & 100.00 \\
     & NBR & +2.25 & 0.3360 & 1.00 & 1.06 & 94.69 & 100.00 & -1.04 & 0.2846 & 1.01 & 1.48 & 94.74 & 100.00 \\
     & EBR & +0.85 & 0.3353 & 0.99 & 1.07 & 94.58 & 100.00 & -0.56 & 0.2872 & 1.00 & 1.46 & 94.65 & 100.00 \\
\addlinespace[3pt]
    100 & GEE & +2.41 & 0.2111 & 0.99 & 1.00 & 94.50 & 100.00 & +1.48 & 0.1920 & 0.98 & 1.22 & 94.19 & 100.00 \\
     & RBR & +0.25 & 0.2065 & 0.99 & 1.06 & 94.53 & 100.00 & -0.41 & 0.1885 & 0.98 & 1.27 & 94.30 & 100.00 \\
     & NBR & +1.31 & 0.2083 & 0.99 & 1.04 & 94.63 & 100.00 & -0.62 & 0.1875 & 0.99 & 1.28 & 94.49 & 100.00 \\
     & EBR & +0.36 & 0.2073 & 0.99 & 1.05 & 94.55 & 100.00 & -0.34 & 0.1886 & 0.98 & 1.27 & 94.32 & 100.00 \\
\addlinespace[3pt]
    500 & GEE & +0.49 & 0.0908 & 1.01 & 1.00 & 94.96 & 100.00 & +0.33 & 0.0817 & 1.01 & 1.24 & 95.09 & 100.00 \\
     & RBR & +0.05 & 0.0904 & 1.01 & 1.01 & 95.07 & 100.00 & -0.06 & 0.0814 & 1.01 & 1.25 & 94.97 & 100.00 \\
     & NBR & +0.28 & 0.0906 & 1.01 & 1.01 & 95.04 & 100.00 & -0.07 & 0.0814 & 1.01 & 1.25 & 94.98 & 100.00 \\
     & EBR & +0.05 & 0.0905 & 1.01 & 1.01 & 95.07 & 100.00 & -0.06 & 0.0814 & 1.01 & 1.25 & 94.98 & 100.00 \\
\bottomrule
\end{tabular}%
}
\end{table}

Under complete data (see Table \ref{tab:sim_binary_complete_merged}), all methods converge in essentially all replications. Finite-sample bias of ordinary GEE is considerably more pronounced for the time-varying covariate $x_2$ than for the time-stationary covariate $x_1$, a pattern also documented by \citet{Touloumis2013} for multinomial responses. For both parameters, bias decreases monotonically as $N$ increases and is negligible by $N=500$. Among the proposed estimators, RBR delivers the most consistent bias reduction across both parameters, working structures, and sample sizes. RE values close to $1$ across all configurations confirm that $\wmSigma{M}$ accurately reflects the true sampling variability for RBR. NBR also reduces bias in most settings and yields similar or higher SRE values than RBR under the exchangeable structure. This is consistent with the theoretical observation that condition \eqref{naive_condition} need not require correct specification of the working association structure, as discussed in Web Appendix D. EBR performs comparably to RBR under complete data, with similar SRE values to those of RBR.

The choice of working association structure affects both bias and overall accuracy. The exchangeable structure yields lower bias for $\beta_2$ than working independence at all sample sizes, and its higher SRE values confirm that accounting for the association structure leads to more efficient estimators. For $\beta_1$, the two working structures produce similar SRE values throughout, indicating comparable overall accuracy. RBR achieves consistent bias reduction across both working structures with RE close to $1$ in all cases, confirming its suitability as the default estimator regardless of the working association specification.

Under MCAR monotone dropout (see Table \ref{tab:sim_binary_mcar_merged}), the patterns from the complete-data scenario are largely preserved, but bias and variability increase for all estimators, most noticeably at small sample sizes $N \in \{20, 35\}$. RBR continues to provide meaningful bias reduction. GEE, RBR, and NBR all converge in essentially all replications. EBR becomes numerically unreliable at small sample sizes: convergence rates fall to $93.78\%$ under independence and $91.70\%$ under the exchangeable odds-ratio structure at $N=20$. Among converged replications, the ESE is substantially inflated relative to the other estimators, yielding SRE values as low as $0.37$. These results confirm the recommendation in Section \ref{sec:practical} that EBR should be treated as a fallback rather than a default.

Empirical coverage under $\wmSigma{M}$ is generally close to the nominal $95\%$ level for GEE, RBR, and NBR, though mild undercoverage is visible for $\beta_2$ at $N=20$, particularly under the MCAR scenario. Coverage improves steadily with $N$ for all three estimators. As expected, differences across methods diminish as $N$ increases, and all four estimators perform virtually identically by $N=500$, confirming that the benefits of bias reduction are concentrated when the number of clusters is small to moderate.

\subsection{Poisson Responses}\label{sec:sim-poisson}

Additional simulations for correlated Poisson responses are reported in Supporting Information, Web Appendix F. These simulations assess whether the proposed adjustments remain stable in settings where ordinary GEE already exhibits relatively small finite-sample bias. The gains from bias reduction were less pronounced than in the binary case. Under the working independence model, the competing estimators behaved similarly, while working AR(1) correlation matrices improved efficiency in moderate and large samples but introduced numerical instability in the smallest samples, particularly under missingness.

\begin{table}[H]
\centering
\caption{Binary responses (probit model), MCAR monotone missingness. Results for $\beta_1=0.5$
and $\beta_2=1$ under working odds-ratio structures (independence and exchangeable).
Columns report Bias ($100\times$Bias), empirical standard error (ESE),
relative efficiency (RE), simulated relative efficiency (SRE),
empirical coverage (EC, \%) of nominal 95\% Wald intervals,
and convergence proportion (CP, \%).}
\label{tab:sim_binary_mcar_merged}
\resizebox{\textwidth}{!}{%
\begin{tabular}{@{}
S[table-format=3.0] l
S[table-format=+3.2] S[table-format=1.4] S[table-format=1.2] S[table-format=1.2]
S[table-format=2.2] S[table-format=3.2]
S[table-format=+3.2] S[table-format=1.4] S[table-format=1.2] S[table-format=1.2]
S[table-format=2.2] S[table-format=3.2]@{}}
\toprule
& & \multicolumn{6}{c}{Independence} & \multicolumn{6}{c}{Exchangeable} \\
\cmidrule(lr){3-8}\cmidrule(lr){9-14}
{$N$} & {Method}
& {Bias} & {ESE} & {RE} & {SRE} & {EC} & {CP}
& {Bias} & {ESE} & {RE} & {SRE} & {EC} & {CP} \\
\midrule

\addlinespace[6pt]
\multicolumn{14}{c}{$\beta_1=0.5$}\\
\addlinespace[3pt]
    20 & GEE & +4.18 & 0.6442 & 0.90 & 1.00 & 92.81 & 100.00 & +3.75 & 0.6204 & 0.93 & 1.08 & 93.66 & 100.00 \\
     & RBR & +0.09 & 0.5797 & 0.96 & 1.24 & 93.41 & 99.99 & -0.39 & 0.5622 & 0.99 & 1.32 & 94.22 & 99.87 \\
     & NBR & +1.23 & 0.5830 & 0.96 & 1.23 & 93.47 & 100.00 & -1.51 & 0.5245 & 1.06 & 1.51 & 95.11 & 100.00 \\
     & EBR & +1.91 & 1.0209 & 389.26 & 0.40 & 93.30 & 93.78 & +2.50 & 0.9074 & 1.06 & 0.51 & 93.89 & 91.70 \\
\addlinespace[3pt]
    35 & GEE & +3.99 & 0.4707 & 0.95 & 1.00 & 93.97 & 100.00 & +3.02 & 0.4552 & 0.97 & 1.07 & 94.38 & 100.00 \\
     & RBR & +0.49 & 0.4406 & 0.99 & 1.15 & 94.48 & 100.00 & -0.17 & 0.4278 & 1.00 & 1.22 & 95.04 & 99.99 \\
     & NBR & +1.55 & 0.4492 & 0.98 & 1.10 & 94.43 & 100.00 & -0.99 & 0.4186 & 1.03 & 1.27 & 95.46 & 100.00 \\
     & EBR & +1.77 & 0.5219 & 0.98 & 0.82 & 94.31 & 97.43 & +1.51 & 0.5627 & 1.05 & 0.70 & 94.77 & 95.98 \\
\addlinespace[3pt]
    50 & GEE & +3.35 & 0.3573 & 0.98 & 1.00 & 94.30 & 100.00 & +2.87 & 0.3450 & 1.00 & 1.07 & 94.79 & 100.00 \\
     & RBR & +0.98 & 0.3411 & 1.01 & 1.11 & 94.80 & 100.00 & +0.65 & 0.3299 & 1.03 & 1.18 & 95.16 & 100.00 \\
     & NBR & +1.90 & 0.3464 & 1.00 & 1.07 & 94.65 & 100.00 & +0.39 & 0.3268 & 1.04 & 1.21 & 95.40 & 100.00 \\
     & EBR & +1.43 & 0.3518 & 1.00 & 1.04 & 94.59 & 99.40 & +1.28 & 0.3549 & 1.00 & 1.02 & 95.07 & 98.85 \\
\addlinespace[3pt]
    100 & GEE & +1.41 & 0.2414 & 0.98 & 1.00 & 94.31 & 100.00 & +1.26 & 0.2363 & 0.98 & 1.04 & 94.31 & 100.00 \\
     & RBR & +0.13 & 0.2351 & 0.99 & 1.06 & 94.61 & 100.00 & +0.00 & 0.2304 & 0.99 & 1.10 & 94.70 & 100.00 \\
     & NBR & +0.71 & 0.2376 & 0.99 & 1.04 & 94.55 & 100.00 & -0.11 & 0.2290 & 1.00 & 1.11 & 94.85 & 100.00 \\
     & EBR & +0.30 & 0.2393 & 0.98 & 1.02 & 94.52 & 99.77 & +0.35 & 0.2354 & 0.98 & 1.05 & 94.57 & 99.04 \\
\addlinespace[3pt]
    500 & GEE & +0.38 & 0.1057 & 1.00 & 1.00 & 94.75 & 100.00 & +0.34 & 0.1036 & 1.00 & 1.04 & 94.82 & 100.00 \\
     & RBR & +0.09 & 0.1051 & 1.00 & 1.01 & 94.81 & 100.00 & +0.05 & 0.1030 & 1.00 & 1.06 & 95.04 & 100.00 \\
     & NBR & +0.23 & 0.1054 & 1.00 & 1.01 & 94.78 & 100.00 & +0.05 & 0.1029 & 1.00 & 1.06 & 95.02 & 100.00 \\
     & EBR & +0.12 & 0.1052 & 1.00 & 1.01 & 94.75 & 100.00 & +0.13 & 0.1031 & 1.00 & 1.05 & 95.03 & 100.00 \\
\addlinespace[3pt]

\addlinespace[6pt]
\multicolumn{14}{c}{$\beta_2=1$}\\
\addlinespace[3pt]
    20 & GEE & +15.01 & 0.6614 & 0.90 & 1.00 & 92.60 & 100.00 & +10.34 & 0.5834 & 0.91 & 1.31 & 92.84 & 100.00 \\
     & RBR & +3.83 & 0.5914 & 0.94 & 1.31 & 93.06 & 99.99 & +0.01 & 0.5279 & 0.95 & 1.65 & 92.99 & 99.87 \\
     & NBR & +5.78 & 0.5859 & 0.96 & 1.33 & 93.39 & 100.00 & -3.55 & 0.4826 & 1.02 & 1.96 & 94.10 & 100.00 \\
     & EBR & +11.20 & 1.1086 & 250.88 & 0.37 & 92.55 & 93.78 & +6.55 & 0.9520 & 2.05 & 0.51 & 92.69 & 91.70 \\
\addlinespace[3pt]
    35 & GEE & +8.18 & 0.4631 & 0.94 & 1.00 & 93.61 & 100.00 & +5.14 & 0.4035 & 0.96 & 1.34 & 93.96 & 100.00 \\
     & RBR & +0.84 & 0.4328 & 0.97 & 1.18 & 94.21 & 100.00 & -1.28 & 0.3796 & 0.98 & 1.53 & 94.25 & 99.99 \\
     & NBR & +3.00 & 0.4372 & 0.97 & 1.15 & 94.25 & 100.00 & -2.72 & 0.3678 & 1.01 & 1.63 & 94.66 & 100.00 \\
     & EBR & +3.22 & 0.5527 & 0.93 & 0.72 & 93.84 & 97.43 & +1.40 & 0.5395 & 0.96 & 0.76 & 94.11 & 95.98 \\
\addlinespace[3pt]
    50 & GEE & +5.94 & 0.3911 & 0.98 & 1.00 & 94.35 & 100.00 & +3.60 & 0.3400 & 0.97 & 1.34 & 94.11 & 100.00 \\
     & RBR & +0.78 & 0.3721 & 1.00 & 1.13 & 94.78 & 100.00 & -0.97 & 0.3247 & 1.00 & 1.48 & 94.38 & 100.00 \\
     & NBR & +2.64 & 0.3768 & 1.00 & 1.10 & 94.71 & 100.00 & -1.37 & 0.3212 & 1.01 & 1.51 & 94.51 & 100.00 \\
     & EBR & +2.00 & 0.4249 & 0.92 & 0.86 & 94.48 & 99.40 & +0.11 & 0.3573 & 0.95 & 1.23 & 94.38 & 98.85 \\
\addlinespace[3pt]
    100 & GEE & +2.97 & 0.2366 & 0.98 & 1.00 & 94.26 & 100.00 & +1.89 & 0.2142 & 0.98 & 1.23 & 94.22 & 100.00 \\
     & RBR & +0.06 & 0.2299 & 0.99 & 1.08 & 94.55 & 100.00 & -0.75 & 0.2088 & 0.98 & 1.30 & 94.17 & 100.00 \\
     & NBR & +1.32 & 0.2321 & 0.99 & 1.05 & 94.62 & 100.00 & -0.96 & 0.2073 & 0.99 & 1.32 & 94.22 & 100.00 \\
     & EBR & +0.53 & 0.2359 & 0.97 & 1.02 & 94.51 & 99.77 & -0.05 & 0.2199 & 0.95 & 1.18 & 94.15 & 99.04 \\
\addlinespace[3pt]
    500 & GEE & +0.55 & 0.1026 & 1.00 & 1.00 & 95.22 & 100.00 & +0.37 & 0.0919 & 1.01 & 1.25 & 94.96 & 100.00 \\
     & RBR & -0.05 & 0.1020 & 1.01 & 1.02 & 95.15 & 100.00 & -0.17 & 0.0913 & 1.01 & 1.26 & 94.99 & 100.00 \\
     & NBR & +0.23 & 0.1022 & 1.01 & 1.01 & 95.13 & 100.00 & -0.17 & 0.0913 & 1.01 & 1.27 & 95.04 & 100.00 \\
     & EBR & +0.00 & 0.1022 & 1.00 & 1.01 & 95.11 & 100.00 & -0.05 & 0.0915 & 1.01 & 1.26 & 95.04 & 100.00 \\
\bottomrule
\end{tabular}%
}
\end{table}

\section{Data Application: Shoulder-tip Pain Trial}\label{sec:data}
\citet{Lumley1996} described a randomized trial evaluating whether abdominal suction reduces shoulder-tip pain following laparoscopic surgery. Forty-one patients were randomized, 22 assigned to abdominal suction and 19 to no suction. Pain was recorded morning and afternoon for three postoperative days ($j=1,\ldots,6$) using a visual analog scale. To focus on the bias-reduction and bias-correction methods of Section \ref{sec:or} for correlated binary responses, we dichotomized pain as low (1) versus high (0).

Let $\pi_{ij}$ denote the probability that patient $i$ reports low pain at time $j$. We fitted the marginal logistic regression model
\[
\log\left( \frac{\pi_{ij}}{1-\pi_{ij}} \right)
=
\beta_0 + \beta_1 x_{1i} + \beta_2 x_{2i} + \beta_3 x_{3i} +
\beta_4 x_{4ij},
\]
where $x_{1i}$ indicates abdominal suction, $x_{2i}$ is baseline age (years), $x_{3i}$ indicates female sex, and $x_{4ij}=I(j\ge 5)$ indicates the final postoperative day ($j=5,6$). Within-subject association was modeled using marginalized odds ratios \citep{Touloumis2026}. The estimated marginalized odds ratios ranged from 3.99 to 135.00, reflecting heterogeneous dependence, so we adopted an unstructured working odds-ratio structure for modeling the association. We report Wald inference based on $\wmSigma{M}$ and focus on GEE, RBR, and RBC in Table \ref{shoulder_or_results_main}, where RBR serves as the representative bias-reducing estimator. Complete results for all estimators are reported in Web Appendix G.

\begin{table}[H]
\centering
\caption{Cholecystectomy (shoulder-tip pain) data: estimates, small-sample adjusted standard errors, and Wald $p$-values for GEE and the robust bias-reduced (RBR) and robust bias-corrected (RBC) estimators under an unstructured working odds-ratio parameterization.}
\label{shoulder_or_results_main}
\resizebox{\textwidth}{!}{%
\begin{tabular}{l
S[table-format=1.4] S[table-format=1.4] S[table-format=1.4]
S[table-format=1.4] S[table-format=1.4] S[table-format=1.4]
S[table-format=1.4] S[table-format=1.4] S[table-format=1.4]}
\toprule
& \multicolumn{3}{c}{GEE} & \multicolumn{3}{c}{RBR} & \multicolumn{3}{c}{RBC} \\
\cmidrule(lr){2-4}\cmidrule(lr){5-7}\cmidrule(lr){8-10}
Parameter
& {Estimate} & {SE} & {$p$-value}
& {Estimate} & {SE} & {$p$-value}
& {Estimate} & {SE} & {$p$-value} \\
\midrule
$\beta_0$ & -2.1898 & 1.0698 & 0.0407 & -2.0360 & 1.0233 & 0.0466 & -2.0118 & 1.0188 & 0.0483 \\
$\beta_1$ &  1.8602 & 0.5794 & 0.0013 &  1.7602 & 0.5566 & 0.0016 &  1.7583 & 0.5549 & 0.0015 \\
$\beta_2$ &  0.0317 & 0.0165 & 0.0548 &  0.0295 & 0.0159 & 0.0626 &  0.0291 & 0.0158 & 0.0654 \\
$\beta_3$ &  0.3478 & 0.5953 & 0.5591 &  0.3065 & 0.5773 & 0.5955 &  0.2925 & 0.5760 & 0.6116 \\
$\beta_4$ &  0.9513 & 0.3894 & 0.0146 &  0.9029 & 0.3770 & 0.0166 &  0.9079 & 0.3766 & 0.0159 \\
\bottomrule
\end{tabular}
}
\end{table}

Inferential conclusions were stable across estimators. Under RBR, both the treatment and final-day indicators were statistically significant at the 5\% level, age was marginally significant at the 10\% level, and there was no evidence of an association with sex. Patients assigned to abdominal suction had higher odds of reporting low pain ($\exp(\widehat{\beta}_1)=5.81$, $95\%$ CI $[2.16,15.65]$), and observations on the final postoperative day were associated with higher odds of low pain relative to earlier assessments ($\exp(\widehat{\beta}_4)=2.47$, $95\%$ CI $[1.25,4.86]$). The treatment effect estimate was $1.76$ under RBR (SE$=0.557$) versus $1.86$ under GEE (SE$=0.579$), a modest reduction in both the point estimate and standard error.

Fitting the same marginal model with an unstructured working correlation matrix produced numerical difficulties, including nonconvergence, inadmissible correlation estimates, and a working correlation matrix that was not positive definite. This is consistent with the known constraints of correlation-coefficient parameterizations for correlated binary responses and supports the use of the odds-ratio parameterization when flexible within-subject association is needed.

\section{Summary and Future Directions}\label{sec:discussion}

Viewing GEE as an $M$-estimation problem provides a unified route to bias reduction and bias correction through adjusted estimating equations. Within this framework, we derived an explicit condition for bias reduction and used it to construct three estimators, robust, naive, and empirical, together with one-step bias-corrected counterparts. All procedures share the first-order asymptotic distribution of ordinary GEE but differ in how the ideal adjustment is approximated and, consequently, in their robustness and efficiency properties. The construction is valid for correlation-coefficient parameterizations and extends naturally to odds-ratio parameterizations for correlated binary responses. It also clarifies how several existing proposals fit within the framework. Under correlation-coefficient parameterizations, the bias-corrected estimator of \citet{Lunardon2017} coincides with our RBC estimator, while the naive bias-reduced and bias-corrected estimators of \citet{Paul_Small_2014} arise as special cases of our NBR and NBC estimators when the working covariance is treated as fixed with respect to the mean parameters. Under this simplification, first-order bias reduction is not guaranteed for mean-dependent variance functions, regardless of whether the working assumption for the association structure is correctly specified, a limitation that the robust and empirical estimators of the present framework avoid.

Across the simulation designs considered, the robust bias-reduced (RBR) estimator retained the large-sample properties of ordinary GEE while achieving smaller finite-sample bias and comparable or improved mean squared error. The benefits were clearest for correlated binary responses under odds-ratio parameterizations, where ordinary GEE exhibited appreciable small-sample bias. The naive estimator can be attractive when the working association is plausibly correct, whereas the empirical estimator tracked RBR closely in complete data but proved less stable in the smallest samples under missingness for correlated binary responses. These results support RBR as a practical default when sample sizes are small to moderate. For Wald-type inference, we recommend the small-sample adjusted covariance estimator $\wmSigma{M}$, which provided reliable calibration across settings. Software implementing all procedures is available in the \texttt{geer} package.

Several directions for further work are suggested. First, adjusted estimating equations motivate a reexamination of model-selection and goodness-of-fit criteria for correlated data, including QIC, QICu, and CIC \citep{pan2001,cui2007,hin2009,hardin2002}, whose standard derivations rely on unadjusted estimating equations. Second, extending the framework to ordinal and nominal multinomial responses would broaden applicability, though the derivatives required for the adjustment terms may be challenging to obtain in closed form when association is parameterized through local odds ratios \citep{Touloumis2013}. Finally, combining bias reduction with the existence guarantees of the penalized GEE framework of \citet{Touloumis2026} within a single approach remains an open question. The main obstacle is that the ideal bias-reduction adjustment depends on the true covariance matrices of the responses, which are unknown and must be estimated, whereas the PGEE penalty depends only on the model-based covariance matrix $\mSigma{0}$ and its derivatives. Reconciling these two adjustments without sacrificing the finiteness guarantees of the penalized framework remains a worthwhile direction for future research.

\bibliographystyle{plainnat}
\bibliography{bibliography}

\includepdf[pages=-]{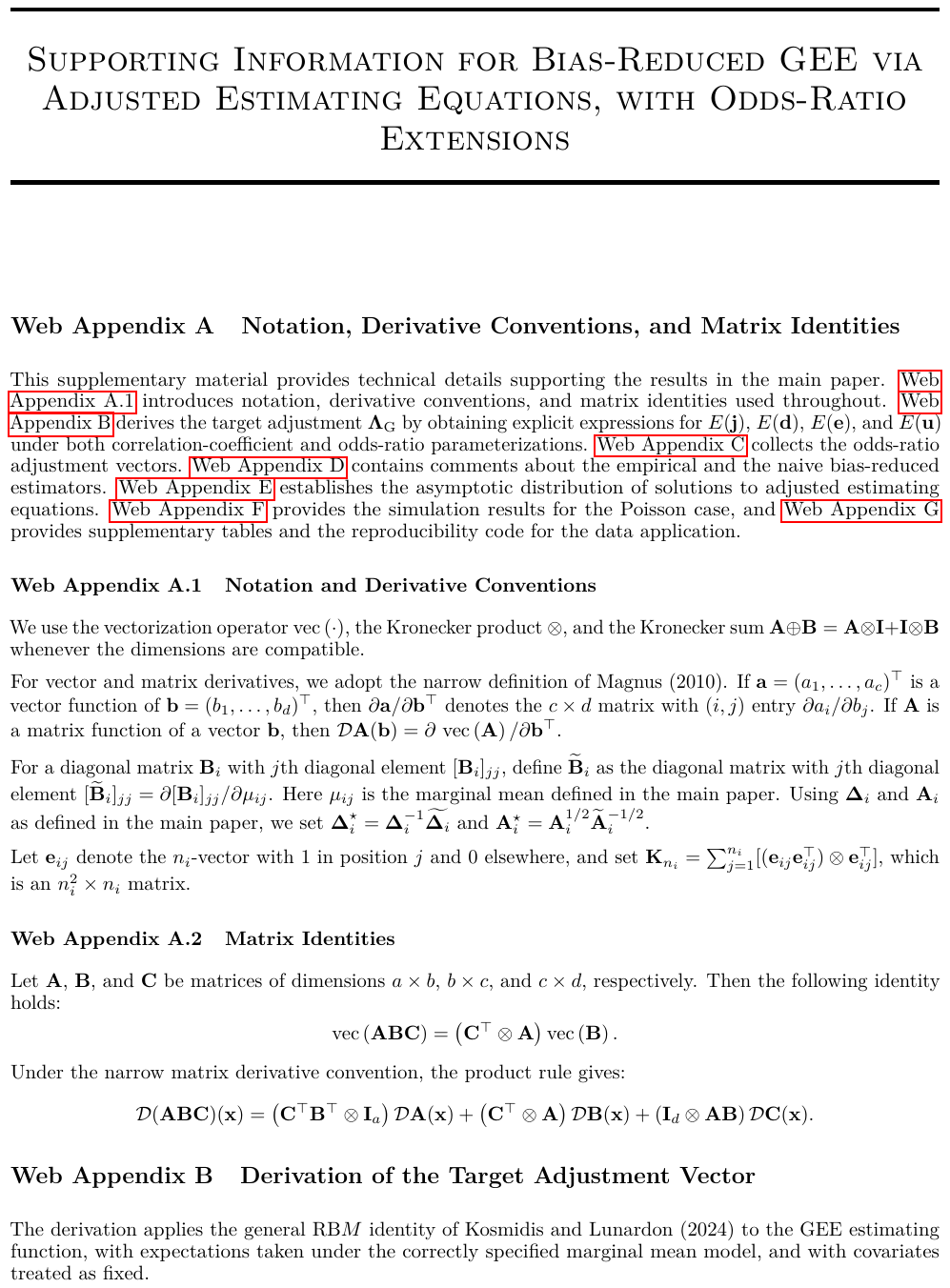}

\end{document}